\documentclass[preprint,aps,showpacs,nofootinbib,preprintnumbers,amsmath,amssymb]{revtex4-1}
\usepackage{amssymb}
\usepackage{epsfig}
\usepackage{graphicx}
\usepackage{subfigure}
\usepackage{dcolumn}
\usepackage{bm}
\usepackage[usenames ,dvipsnames]{xcolor}
\usepackage{slashed}

\begin{document}

\title{Multicomponent Dark Matter in the Light of CALET and DAMPE}

\author{Chao-Qiang~Geng$^{1,2,3}$\footnote{geng@phys.nthu.edu.tw},
Da~Huang$^{4,5,6}$\footnote{dahuang@bao.ac.cn}, and Lu Yin$^{2}$\footnote{yinlu@gapp.nthu.edu.tw}
}
 \affiliation{$^{1}$Chongqing University of Posts \& Telecommunications, Chongqing 400065\\
  $^{2}$Department of Physics, National Tsing Hua University, Hsinchu 300\\
  $^{3}$Physics Division, National Center for Theoretical Sciences, Hsinchu 300 \\
  $^{4}$National Astronomical Observatories, Chinese Academy of Sciences, Beijing, 100012 \\
  $^{5}$Faculty of Physics, University of Warsaw, Pasteura 5, 02-093 Warsaw  \\
  $^{6}$Departamento de Fisica da Universidade de Aveiro and CIDMA, Campus de Santiago, 3810-183 Aveiro, Portugal}

\date{\today}
\begin{abstract}
In the light of the latest measurements on the total $e^+ + e^-$ flux by CALET and DAMPE experiments, we revisit the multicomponent leptonically decaying dark matter (DM) explanations to the cosmic-ray electron/positron excesses observed previously. Especially, we  use the single and double-component DM models to explore the compatibility of the AMS-02 positron fraction with the new CALET or DAMPE data. 
As a result, for either the AMS-02 + CALET dataset or the AMS-02 + DAMPE one, both the single and double-component DM models can provide reasonable fits. If we further take into the diffuse $\gamma$-ray constraints from Fermi-LAT, only the double-component DM models are allowed.
\end{abstract}

\maketitle

\section{Introduction}
\label{s1}
High-energy cosmic ray (CR) particles, including electrons, positrons, (anti)protons, and various nuclei, are important targets for the dark matter (DM) indirect searches~\cite{Bertone:2004pz,Feng:2010gw}, since they might be the products of DM annihilations/decays. Recently, CALorimetric Electron Telescope (CALET) has published a new measurement on the total $e^+ + e^-$ flux spectrum in~\cite{CALET, CALET19}, which extended the energy range compared with their 2017 data~\cite{CALET17}. In comparison, DArk Matter Particle Explorer (DAMPE)~\cite{DAMPE} provided a similar measurement on the same spectrum with the highest energy at 4.6~TeV. These new measurements provide us several new features of the CR electrons/positrons. The DAMPE data~\cite{DAMPE} can be well fitted by a smooth broken-power-law spectrum and displays a spectral break at about 0.9~TeV. It also exhibits an intriguing peak-like excess around 1.4~TeV. On the other hand, the latest CALET data~\cite{CALET} agrees with other experiments up to several tens of GeV, while it is lower than those of DAMPE from 30~GeV to 800 GeV. It also confirms a flux suppression above 1 TeV suggested by the DAMPE data. However, it does not show the 1.4~TeV peak.

One remarkable discovery in the DM indirect searches is the observations of the possible electron/positron excesses by PAMELA~\cite{PAMELA,PAMELA2}, Fermi-LAT~\cite{FermiLAT,FermiLAT1, FermiLATp}, AMS-02~\cite{AMS02, AMSf, AMSep, AMSt, AMSp19}, and many others~\cite{ATIC,AMS01,HESS, HESS1,Archer:2018chh}. Such excesses indicate that there are some unknown nearby $e^+/e^-$ sources in our Galaxy. One kind of popular candidates is provided by the DM annihilations~\cite{DMindependent, annihilation, AnnihilationDecay, Jin:2014ica,Lin:2014vja,Boudaud:2014dta,Yuan:2017ozr} or decays~\cite{AnnihilationDecay, Jin:2014ica,Lin:2014vja, decay, Ishiwata:2009vx, 3bodydecay,2body,2bodyAMSa}, while other explanations involve pulsars~\cite{pulsar} or some exotic astrophysical $e^+/e^-$ acceleration mechanisms~\cite{astrophysical}. Note that the DM scenarios have been stringently constrained by the measured CR antiproton-to-proton ratio and (anti-)proton fluxes~\cite{PAMELApr, PAMELApr1, AMSpr, AMSapr} which agreed with the theoretical astrophysical predictions very well~\cite{protonBound}, in spite of some recent tantalizing hints for the antiproton excesses at high energies~\cite{protonEx}. In the literature, a simple way to evade this constraint is the so-called leptophilic DM scenario in which only the leptons are allowed to couple to the dark sector. However, even in this leptophilic setup, the single-component annihilating/decaying DM models have been shown to be unable to explain all the existing data~\cite{2bodyAMSa}.

In Refs.~\cite{2DM_1, 2DM_2, 2DM_3, 2DM_4, 2DM_5}, the multicomponent leptophilic decaying DM models~\cite{2DM_1, 2DM_2, 2DM_3, 2DM_4, 2DM_5, 2DM_others} have been proposed to interpret this puzzling situation. In Ref.~\cite{2DM_1}, the spectra of the AMS-02 positron fraction and the Fermi-LAT total $e^+ + e^-$ have been demonstrated  to be well fitted by the double-component DM models which could also satisfy the diffuse $\gamma$-ray constraints by Fermi-LAT~\cite{FermiLAT_Gamma}. One bonus of this double-component DM model is that the apparent substructure appearing around 100~GeV can be interpreted as the energy cutoff of the light DM particle decays. In Refs.~\cite{2DM_3, 2DM_4, 2DM_5}, this double-component DM model is also investigated by using the latest AMS-02 $e^+/e^-$ data release and shown to provide a good fit to the data.

In this work, we  explore the implications of the latest CALET~\cite{CALET,CALET19} and DAMPE~\cite{DAMPE}
data to the $e^+/e^-$ excesses observed before. Concretely, we would like to apply our single and double-component DM models~\cite{2DM_1, 2DM_2, 2DM_3, 2DM_4, 2DM_5} to fit the AMS-02 positron fraction and the CALET or DAMPE total $e^+ + e^-$ flux data. Due to the explicit inconsistency between the CALET and DAMPE data~\cite{CALET}, we do not try to combine them. Furthermore, we only take into account the continuous spectrum in the DAMPE data~\cite{Yuan:2017ysv,Niu:2017hqe,DAMPEcont}, and do not make any attempt to explain the peak structure at 1.4~TeV~\cite{DAMPE} which may be explained by the some nearby DM substructures~\cite{DAMPEmodels,Yuan:2017ysv,Huang:2017egk}. Note that more recently AMS-02 has published  the new results on the positron flux, which shows spectral softening around 1 TeV~\cite{AMSp19}. However, in the present paper we still use the earlier positron fraction data in Ref.~\cite{AMSf}, since it describes the positron spectral shape more accurately than the positron flux due to its greatly reduced systematic errors.

The paper is organized as follows. In Sec.~\ref{s2}, we briefly summarize our multicomponent decaying DM framework and the procedure to obtain the electron/positron fluxes. The fitting results with the CALET and DAMPE data are given in Sec.~\ref{s3} and \ref{s4}, respectively. Finally, we conclude and further discuss our results in Sec.~\ref{s5}.

\section{Multi-Component Dark Matter Framework and Galactic Cosmic Ray Propagation}
\label{s2}
Following Refs.~\cite{2DM_1, 2DM_2, 2DM_3,2DM_4, 2DM_5}, the $e^-/e^+$ fluxes in our multicomponent decaying DM models can be expressed as:
\begin{eqnarray}\label{totFlux}
\Phi^{(\rm{tot})}_{e^-} &=& \kappa_1 \Phi^{(\rm{primary})}_{e^-} + \kappa_2 \Phi^{(\rm{secondary})}_{e^-}+\Phi^{\rm{DM}}_{e^-} , \nonumber\\
\Phi^{(\rm{tot})}_{e^+} &=& \kappa_2 \Phi^{(\rm{secondary})}_{e^+} + \Phi^{\rm{DM}}_{e^+}.
\end{eqnarray}
$\Phi^{(\rm primary)}_{e^-}$ represents the primary electron flux, which is assumed to be originated from the supernova remnants~\cite{SNR} in the Milky Way. The primary electron injection spectrum is taken as the broken-power law $q^{e}(\rho) \propto \left(\rho/\rho_{e1,2} \right)^{-\gamma_{e1,2,3}}$, in which $\rho_{e1,2}$ are two reference rigidities and $\gamma_{e1,2,3}$ the spectral indices of the three pieces. $\Phi^{(\rm secondary)}_{e^-,e^+}$ denote the secondary positron/electron flux spectra,
which are produced by the collisions of the CR charged nuclei with the interstellar medium in our Galaxy. The primary and secondary electrons/positrons comprise the background $e^+/e^-$ fluxes in our DM model. In this work, the CR propagation is taken to be the diffusion-reacceleration model. As usual, the spatial diffusion coefficient is given by $D_{xx} = \beta D_0 (\rho/\rho_r)^\delta$, where $\rho_r$ is the reference rigidity, $\beta = v/c$ is the velocity and $\delta$ is its spectral index, respectively. The reacceleration is parametrized by $D_{pp}=4 v_A^2 p^2/(3D_{xx}\delta(4-\delta^2)(4-\delta))$. The primary CR proton spectrum, which is needed to calculate the secondary $e^+/e^-$ components, is also assumed to be a broken-power-law function with $q^n (\rho)\propto (\rho/\rho_n)^{\gamma_{n1,2}}$. Due to the complication of the CR propagation processes, we adopt the public GALPROP code~\cite{GALPROP} to calculate the primary and secondary $e^+/e^-$ fluxes with fixed model parameters given in Table~\ref{parameters}, which are consistent with the most recent AMS-02 (anti-)proton and nuclei data~\cite{diffNew, protonEx}. 
Here, in order to improve our fits, we choose different values of the primary electron spectrum index $\gamma_{e3}$ for the AMS-02+CALET and AMS-02+DAMPE datasets.
Finally, the $e^-/e^+$ fluxes with energy below 10~GeV are affected greatly by the solar modulation, for which we take the simple force-field approximation~\cite{SolarModulation} with the Fisk potential as $\phi_F = 0.55$~GV.
 In Eq.~(\ref{totFlux}), we have inserted two parameters $\kappa_1$ and $\kappa_2$ to take into account the possible normalization uncertainties in the primary and secondary fluxes with their values to be determined by fitting the experimental data.

\begin{table}
\caption{Parameters for the diffuse propagation, primary electrons, and primary protons. }
\begin{tabular}[t]{|cccc|ccccc|ccc|}
\hline
\multicolumn{4}{|c|}{diffuse coefficients}&\multicolumn{5}{c|}{primary electrons}
&\multicolumn{3}{c|}{primary protons} \\
\hline
$D_0(\mathrm{cm}^2\mathrm{s}^{-1})$ & $\rho_r(\rm{GV})$  &$\delta$ &$v_A({\rm km\, s}^{-1})$&$\rho_{e1}(\rm{GV})$& $\rho_{e2}(\rm{GV})$ & $\gamma_{e1} $&$\gamma_{e2}$ & $\gamma_{e3}$
&$\rho_{n}(\rm{GV})$&$\gamma_{n1} $&$\gamma_{n2}$\\
\hline
$7.24\times10^{28}$& $4.0$  & 0.38 & $38.5$
&$3.0$ & 61.2 & 1.50 & 2.97 & {2.66/2.50} & $12.88$ & 1.69 & 2.37 \\
\hline
\end{tabular}\label{parameters}
\end{table}

$\Phi^{\rm DM}_{e^-/e^+}$ represent the electron/positron flux contributions from DM decays. In the multicomponent DM framework, the total DM energy density $\rho({\bf x})$ in our Galaxy is taken by $N$ DM components $\chi_i$ with $i=1,2,\cdots,N$ with the equal density $\rho_i ({\bf x}) = \rho({\bf x})/N$, where $\rho({\bf x})$ is parametrized by the isothermal profile~\cite{isothermal}. All of the DM components are assumed to decay via the following channels
\begin{equation}\label{DecayChannel}
\chi_i \to l^{\pm} Y^{\mp},
\end{equation}
where $l$ is the charged leptons, and $Y$ is a charged particle beyond the Standard Model whose origin and decays are not relevant to our discussion below. The decay processes in Eq.~(\ref{DecayChannel}) are  simple realizations in the leptophilic scenario, which are easily satisfied by the recent constraints from the CR antiproton measurements~\cite{PAMELApr, PAMELApr1, AMSpr, AMSapr,protonBound} and CMB power spectrum~\cite{CMBpower}. The differential $e^-/e^+$ multiplicity for each DM component of $\chi_i$ can be written as the summation of the three leptonic channels:
\begin{eqnarray}\label{Norm_Spectrum}
\Big(\frac{dN_{e,p}}{dE}\Big)_i=\frac{1}{2}\Big[\epsilon^e_i\Big(\frac{dN^e} {dE}\Big)_i+\epsilon^\mu_i\Big(\frac{dN^\mu}{dE}\Big)_i+\epsilon^\tau_i\Big(\frac{dN^\tau}{dE}\Big)_i\Big]\;,
\end{eqnarray}
where $\epsilon^{e,\mu,\tau}_i$ denote the branching ratios for $e,\mu,\tau$ channels and satisfy the condition $\epsilon^e_i + \epsilon^\mu_i + \epsilon^\tau_i =1$. Here, the factor of $1/2$ is the consequence of the fact that $e^+$ and $e^-$ come out from two different channels. It is remarkable that the normalized injection spectra for the three two-body processes in Eq.~(\ref{DecayChannel}) can be fixed by the kinematics. Concretely, for $e$ and $\mu$-channels, the injection spectra can be obtained analytically as
\begin{eqnarray}
\Big( \frac{dN^e}{dE} \Big)_i &=& \frac{1}{E_{ci}}\delta(1-x),\\
\Big( \frac{dN^\mu}{dE} \Big)_i &=& \frac{1}{E_{ci}}[3(1-x^2)-\frac{4}{3}(1-x)]\theta(1-x),
\end{eqnarray}
with $x=E/E_{ci}$, while for the $\tau$-channel, its spectrum can be calculated with PYTHIA~\cite{PYTHIA}. The electron/positron energy cutoff $E_{ci}$ for each DM component can be simply obtained via
\begin{eqnarray}\label{cutoff}
E_{ci} = \frac{M_i^2-M_Y^2}{2M_i}\, ,
\end{eqnarray}
in which $M_{i(Y)}$ represents the mass of the $i$-th DM particle (the particle $Y$). Thus, we can write the $e^{\pm}$ source functions from DM decays in the diffusion equation as
\begin{eqnarray}\label{source}
Q^{\rm DM}_{e,p} ({\bf x}, p) = \sum_i \frac{\rho_i ({\bf x})}{\tau_i M_i} \left(\frac{dN_{e,p}}{dE}\right)_i,
\end{eqnarray}
where $M_i$, $\tau_i$ and $\rho_i(\bf x)$ denote the $i$-th DM particle's mass, lifetime and energy density distribution, respectively.
The charged particles $e^\pm$ produced by DM decays propagate through the Galaxy and, finally, reach the Earth so that we can detect the electron/positron signals $\Phi^{\rm DM}_{e^+/e^-}$. Note that $e^\pm$ propagations are very complex~\cite{DiffuseEquation}, involving the the deflection in the galactic magnetic fields and the energy losses via the bremsstrahlung, synchrotron radiations and inverse Compton scatterings. In order to  consistently predict the final fluxes $\Phi^{\rm DM}_{e^+/e^-}$, we use the GALPROP code~\cite{GALPROP} to precisely solve the propagation processes with the same diffusion coefficients as the background electrons and positrons in Table~\ref{parameters}~\cite{diffNew}.

\section{Fitting Results with CALET}\label{s3}
We firstly consider our single and double-component DM models by fitting the combined dataset of the CALET total $e^+ + e^-$ flux~\cite{CALET} and AMS-02 positron fraction~\cite{AMSf}. Note that the latest release of the $e^++e^-$ flux data by CALET in Ref.~\cite{CALET} contains 40 data points ranging from 11~GeV to 4.8~TeV, while the AMS-02 measurement of the positron fraction in Ref.~\cite{AMSf} has 43 data points above 10~GeV, which are less affected by the solar modulation. Thus, there are totally 83 data points in our simple $\chi^2$ fitting.

\subsection{Single-Component Dark Matter}\label{sCALET}
In this subsection, we focus on the single-component DM models. Note that the fitting can only give the best-fit value for the combination $m_{\rm DM} \tau_{\rm DM}$. Thus, without loss of generality, the DM lifetime is meaningful only when the DM mass is determined. We choose several typical lepton energy cutoffs to be 600, 800, 1000, 1200, and 1500~GeV, and the corresponding DM masses $m_{\rm DM}$ can be fixed by kinematics via Eq.~(\ref{cutoff}) with the mass of $Y$ being $m_Y=300~{\rm GeV}$. We further allow  DM to decay
through  $e$, $\mu$ and $\tau$-channels simultaneously and ask for the respective branching ratios $\epsilon_{e,\mu,\tau}$ with the normalization constraint $\epsilon_e + \epsilon_\mu + \epsilon_\tau=1$. Together with  $\kappa_1$ and $\kappa_2$ to account for the primary and secondary $e^-/e^+$ uncertainties, there are totally 6 parameters in the present single-component DM models.

\begin{table}[ht]
\caption{Parameters leading to the minimal values of $\chi^2$ with the single DM cutoffs being 600, 800, 1000, 1200 and 1500~GeV, respectively.}
{\begin{tabular}{@{}cc|cc|ccc|c|cc@{}} \toprule
$E_{c}({\rm GeV})$ & $m_{\rm DM} ({\rm GeV})$ & $\kappa_1$ & $\kappa_2$ & $\epsilon^{e}$& $\epsilon^{\mu}$  & $\epsilon^{\tau}$
 & $\tau(10^{26}{\rm s})$ & $\chi^2_{\rm min}$ & $\chi^2_{\rm min}/{\rm d.o.f.}$
\\
\colrule
600 & 1271 & 0.95  & 1.62  &  0.09 & 0 & 0.91  & 1.54 & 90.4 & 1.17 \\
800 & 1654 & 0.94 & 1.64  & 0.05  & 0 & 0.95 & 1.29 & 89.0 & 1.16 \\
1000 & 2044 & 0.94 & 1.65  & 0.03 & 0 & 0.97 & 1.13 & 91.3 & 1.19 \\
1200 & 2437 & 0.94 & 1.66 & 0.01 & 0 & 0.99 & 1.01 & 97.6 & 1.27 \\
1500 & 3030 & 0.93 & 1.66 & 0 & 0 & 1.0 & $0.88$ & 116.1 & 1.51 \\
\botrule
\end{tabular}\label{tab1DM_CALET}}
\end{table}

\begin{figure}[ht]
\centering
\includegraphics[width=0.34\textwidth, angle =-90]{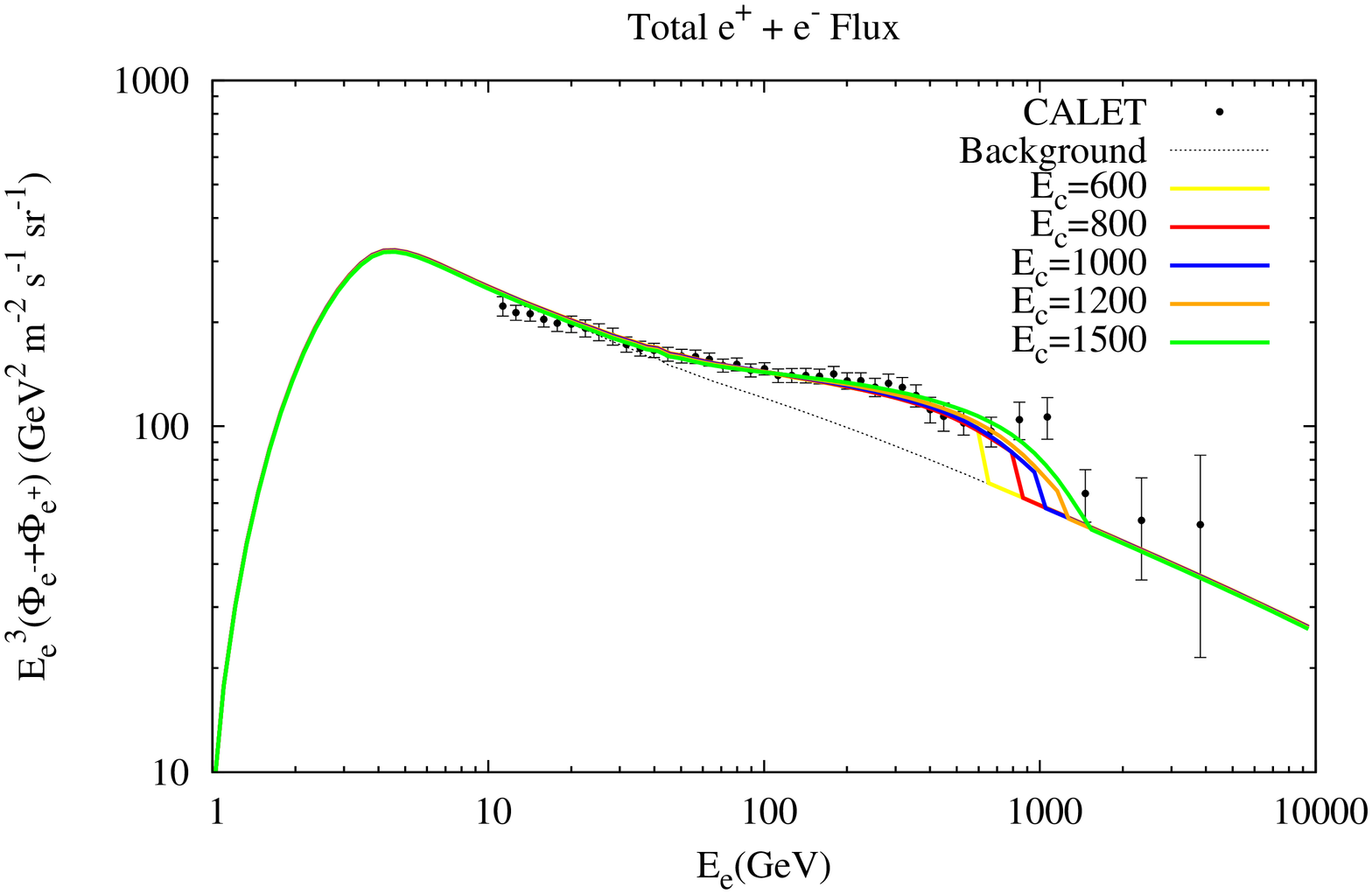}
\includegraphics[width=0.34\textwidth, angle =-90]{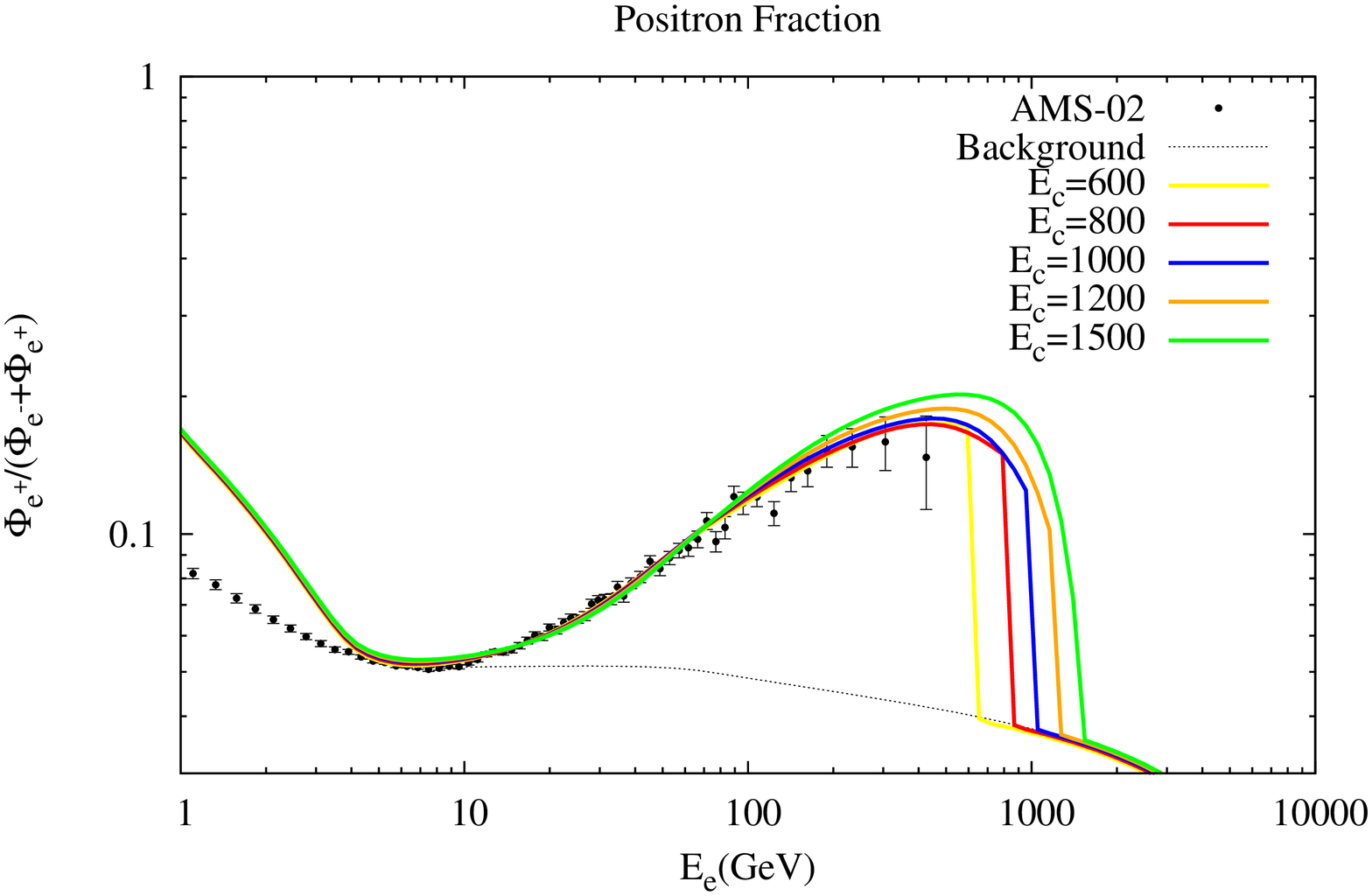}
\caption{The best-fit total $e^+ + e^-$ flux (left panel) and positron fraction (right panel) for single-component DM models with the data from the CALET $e^+ + e^-$ flux and the AMS-02 positron fraction. In each plot, the solid yellow, red, blue, orange and green lines (from left to right) correspond to the cases with the DM electron cutoffs to be $E_{c}=600$, 800, 1000, 1200 and 1500 GeV, respectively, while the dotted black line represents the common background $e^+/e^-$ spectra.
}\label{fig1DM_CALET}
\end{figure}

The fitting results are presented in Table~\ref{tab1DM_CALET} and  Fig.~\ref{fig1DM_CALET}.
In Table~\ref{tab1DM_CALET},
it is shown  that the single-component DM model with the energy cutoff smaller than 1~TeV can provide a relatively good fit to the combined CALET and AMS-02 data as the values of $\chi^2_{\rm min}/{\rm d.o.f}$ are all around 1. Moreover,  $\chi^2_{\rm min}$ has a minimum when $E_c = 800$~GeV, indicating that the combined data favors a decaying DM particle with its mass around 1.6~TeV. As for the lepton flavor dependence, the $\tau$-channel overwhelms the other two leptonic final states. However, it has been found~\cite{Ishiwata:2009vx,Cirelli:2012ut,Papucci:2009gd,Beacom:2004pe, Gamma_DM, Gamma_extraDMIC, Gamma_extraDMIC1, Gamma_extra, Ibarra:2007wg, Lipari:2018gzn} that such a $\tau$-favored heavy DM particle with $m_{\rm DM}\gtrsim 500$~GeV is ruled out by the constraints from the diffuse $\gamma$-ray observations by Fermi-LAT, which suggests that we need to consider some extensions of this simple single-component DM model.

\subsection{Double-Component Dark Matter}\label{dCALET}
Motivated by the substructure occurred at around 100~GeV in most $e^+/e^-$ data,
we have proposed the two-component DM model in which such a substructure could be interpreted as a light DM particle with its lepton cutoff at this energy scale. In this section, we investigate this model with the latest CALET and AMS-02 data. With the particle $Y$ mass of $m_Y = 300$~GeV as the single-component DM models, the light DM mass can be fixed to be $m_L=416$~GeV for $E_{cL} = 100$~GeV, while the heavy DM cutoffs are chosen to be $E_{cH} = 600$, 800, 1000, 1200 and 1500~GeV with its mass $m_H$ determined according to Eq.~(\ref{cutoff}). As mentioned before, due to the strong constraints from the Fermi-LAT measurement of the diffuse $\gamma$-ray~\cite{Ishiwata:2009vx,Cirelli:2012ut,Papucci:2009gd,Beacom:2004pe, Gamma_DM, Gamma_extraDMIC, Gamma_extraDMIC1, Gamma_extra, Ibarra:2007wg,Lipari:2018gzn}, the heavy DM is only allowed to decay into $e$ and $\mu$-leptons, while there is not such a constraint for the light one. In sum, there are 9 free parameters in the present two-component DM models.

\begin{table}[ht]
\caption{Parameters leading to the minimal values of $\chi^2$ with the cutoffs of the heavy DM being 600, 800, 1000, and 1500 GeV, respectively.}
{\begin{tabular}{@{}cc|cc|c|c|cc|cc@{}} \toprule
$E_{cH}({\rm GeV})$& $m_H({\rm GeV})$ & $\kappa_1$ & $\kappa_2$ & $\epsilon^{e,\mu,\tau}_{L}$& $\epsilon^{e,\mu}_{H}$  & $\tau_{L}(10^{26}{\rm s})$
 & $\tau_{H}(10^{26}{\rm s})$ & $\chi^2_{\rm min}$ & $\chi^2_{\rm min}/{\rm d.o.f.}$
\\
\colrule
600 & 1271 & 0.95 & 1.52 & 0.01 ,0.02, 0.97 & 0.20, 0.80 & 1.46 & 1.61 & 81.7 & 1.10 \\
800 & 1654 & 0.94 & 1.52 & 0.03, 0.07, 0.90 & 0.10, 0.90 & 1.51 & 1.34 & 73.2 & 0.99 \\
1000 & 2044 & 0.94 & 1.53 & 0.06, 0.18, 0.76 & 0.05, 0.95 & 1.76 &  1.18 & 70.7 & 0.96\\
1200 & 2437 & 0.94 & 1.56 & 0.10, 0.40, 0.50 & 0.05, 0.95 & 2.34 & 1.31 & 72.5 & 0.98 \\
1500 & 3030 & 0.94 & 1.58 & 0.18, 0.82, 0.00  & 0, 1 &  3.43 & 0.96 & 82.1 & 1.11\\
\botrule
\end{tabular}\label{tab2DM_CALET}}
\end{table}

\begin{figure}
\centering
\includegraphics[width=0.34\textwidth, angle =-90]{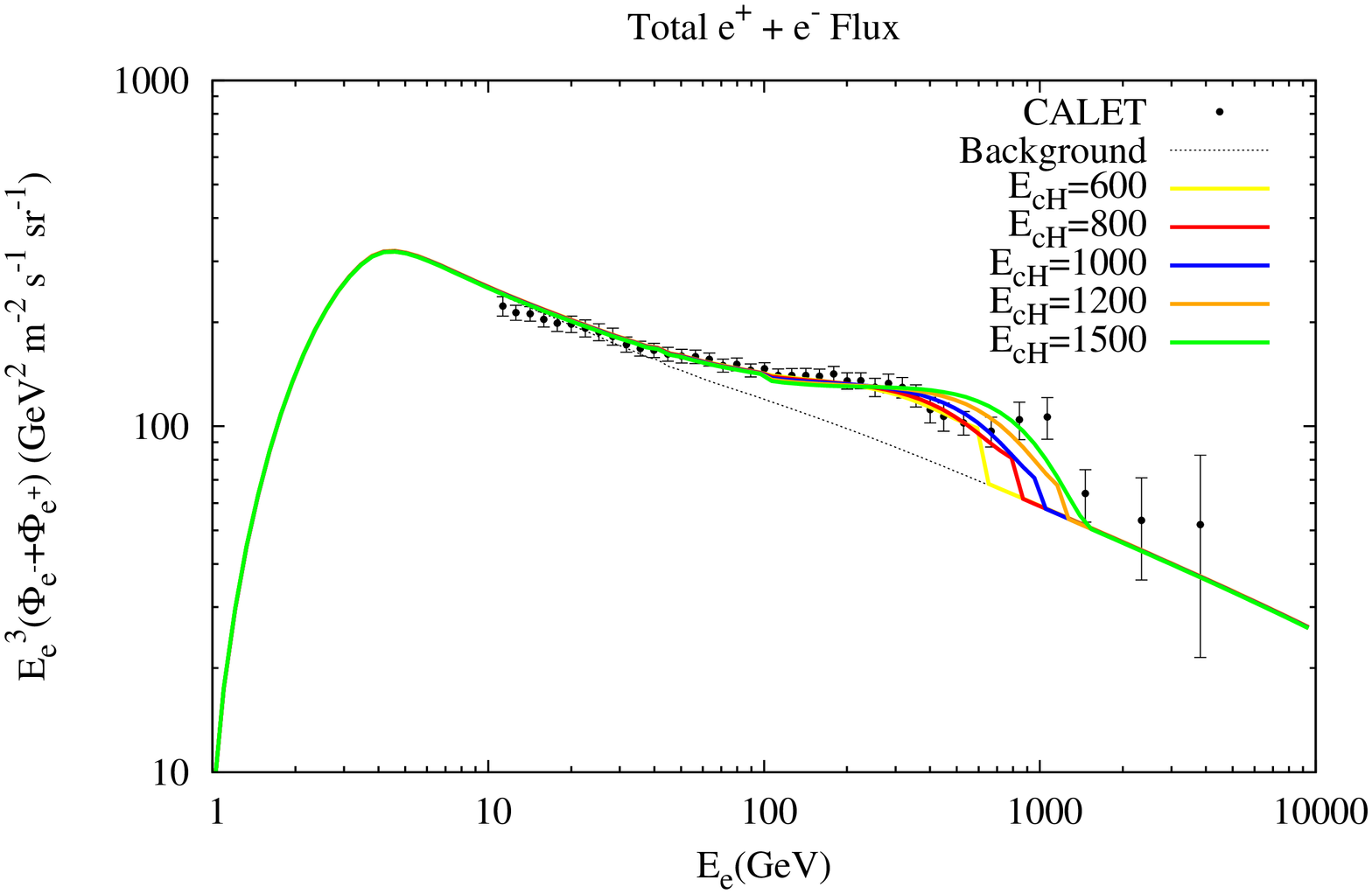}
\includegraphics[width=0.34\textwidth, angle =-90]{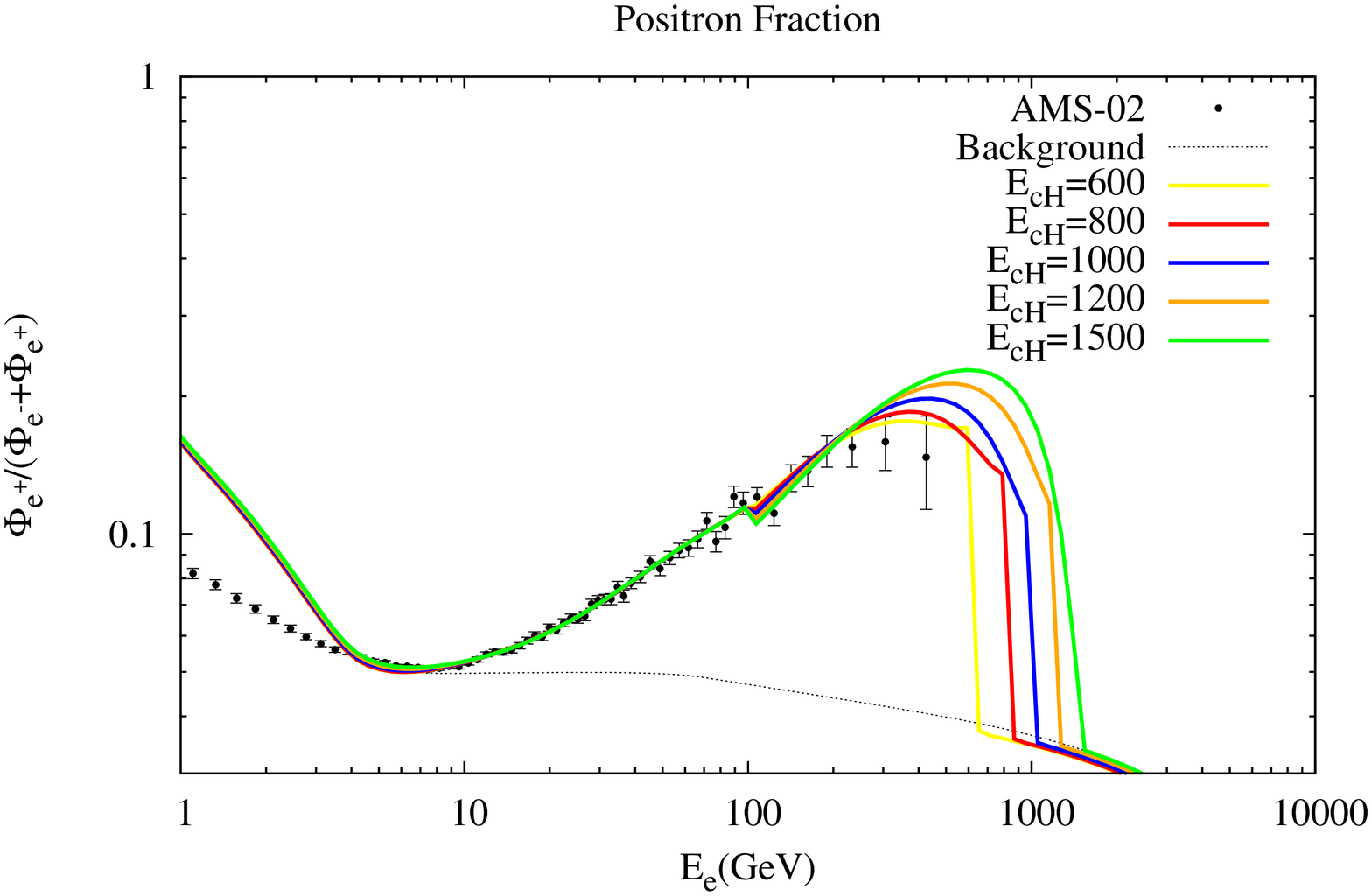}
\caption{The best-fit total $e^+ + e^-$ flux (left panel) and positron fraction (right panel) for double-component DM models with the data from the CALET total $e^+ + e^-$ flux and the AMS-02 positron fraction. In each plot, the solid yellow, red, blue, orange and green lines (from left to right) correspond to the cases with the heavy DM electron cutoffs to be $E_{cH}=600$, 800, 1000, 1200 and 1500 GeV, respectively, while the dotted black line represents the common background $e^+/e^-$ spectra.
}\label{fig2DM_CALET}
\end{figure}

Table~\ref{tab2DM_CALET} and Fig.~\ref{fig2DM_CALET} summarize the final fitting results for the two-component DM models. It is clear from the values of $\chi^2_{\rm min}/{\rm d.o.f.}$ that this two-component DM framework can give a reasonable fit to the AMS-02 and CALET data simultaneously for ${\cal O}(1~{\rm TeV})$ heavy DM cutoffs. It is interesting to note that the minimum value of
$\chi^2_{\rm min}=0.96$ can be obtained with the heavy DM cutoff around 1~TeV. For all cases in Table~\ref{tab2DM_CALET}, the heavy DM decay is always dominated by the $\mu$-channel, while for the light DM the most favored channel transits from $\tau$ for $E_{cH} < 1$~TeV to $\mu$ for $E_{cH} \gtrsim 1.5$~TeV.

\section{Fitting Results with DAMPE}\label{s4}
In this section, we adopt the latest AMS-02 positron fraction~\cite{AMSf} and the new DAMPE total $e^+ + e^-$ flux~\cite{DAMPE} as our dataset for our fitting to the multi-component DM models. Since the CR electron/positron fluxes below 10~GeV suffer from large solar modulation effects, we only use the AMS-02 data above this energy threshold, so that there are totally 43 data points. For the DAMPE data, the lowest energy is 24~GeV, so that all of the 38 points should be considered. In sum, 81 data points will be used in our fittings. Below we discuss the single and double-component decaying DM models, respectively. In the present work, we focus on the DM interpretation of the continuous spectrum in the DAMPE data~\cite{DAMPE,Yuan:2017ysv}, rather than the peak at 1.4 TeV. The latter phenomenon might be the sign of some local structures near our solar system~\cite{DAMPEmodels,Yuan:2017ysv,Huang:2017egk}, which is beyond the scope of the present study. 

\subsection{Single-Component Dark Matter}
\begin{figure}[ht]
\centering
\includegraphics[width=0.34\textwidth, angle =-90]{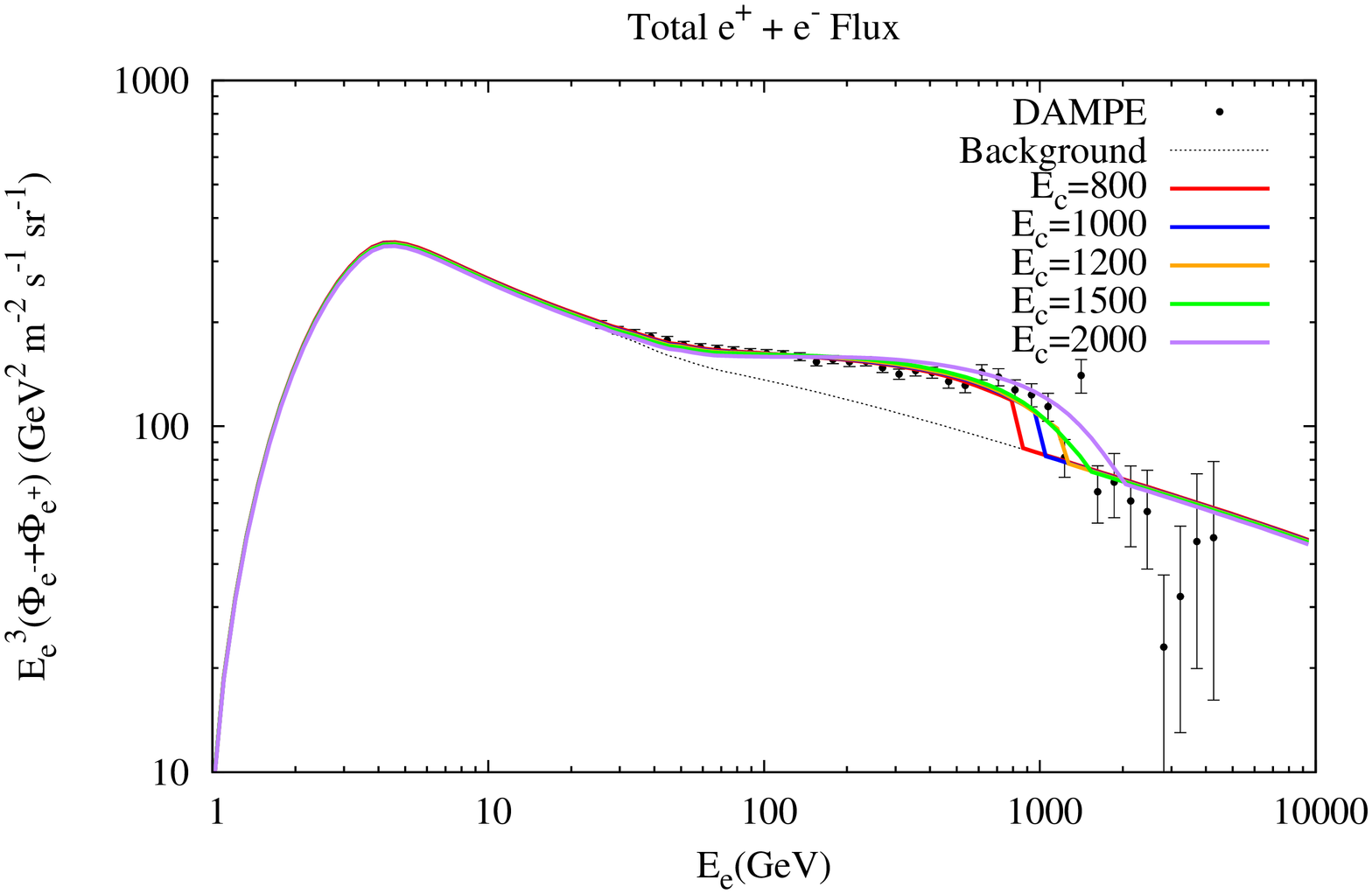}
\includegraphics[width=0.34\textwidth, angle =-90]{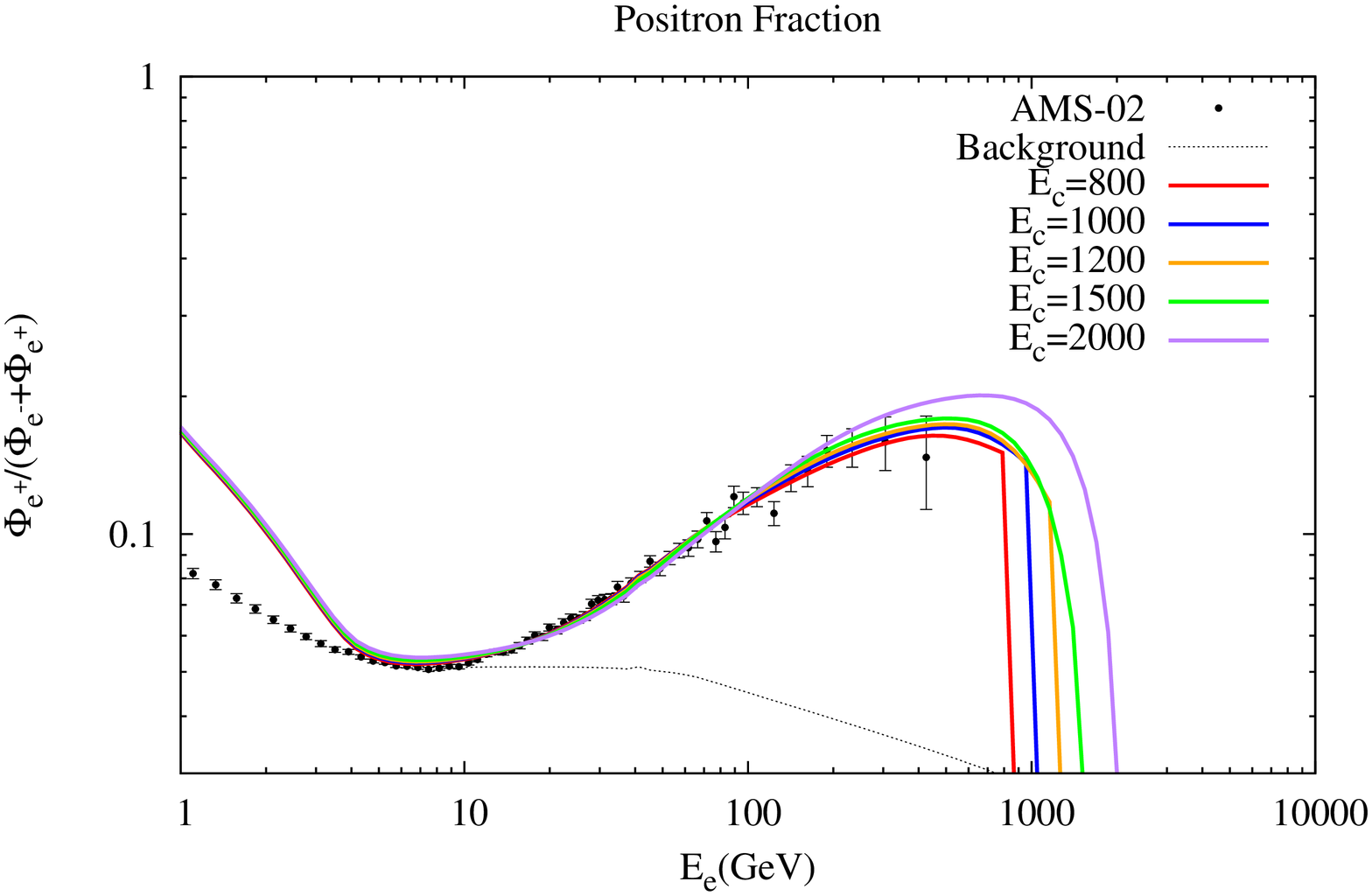}
\caption{The best-fit total $e^+ + e^-$ flux (left panel) and positron fraction (right panel) for single-component DM models with the data from the DAMPE $e^+ + e^-$ flux and the AMS-02 positron fraction. In each plot, the solid red, blue, orange, green and purple lines (from left to right) correspond to the cases with the DM electron cutoffs to be $E_{c}=800$, 1000, 1200, 1500 and 2000 GeV, respectively, while the dotted black line represents the common background $e^+/e^-$ spectra.
}\label{Fig1DM_DAMPE}
\end{figure}

Following the same procedure in Sec.~\ref{sCALET}, we obtain the best-fit parameters and the minimum $\chi^2$ as shown in Table~\ref{tab1DM_DAMPE} for the single DM cutoff chosen as $E_c = 800$, 1000, 1200, 1500 and 2000~GeV, respectively. The corresponding best-fit spectra for the total $e^+ + e^-$ flux and the positron fraction are illustrated in Fig.~\ref{Fig1DM_DAMPE}. 
 In our fitting, we find that the resultant $\chi^2$ of each case is sensitive to the primary electron background spectrum, especially the value of $\gamma_{e3}$. Consequently, the fit can be improved a lot if we take a little smaller values of $\gamma_{e3}$ than that in the fit with the CALET+AMS-02 dataset. Therefore, we would like to fix $\gamma_{e3} = 2.5$ in the rest of this section. It is clear from Table~\ref{tab1DM_DAMPE} that, except the case with $E_c = 2000$~GeV, all of the single-component DM models with different $E_c$ can give reasonable fittings to the DAMPE+AMS-02 data, as the reduced $\chi^2$ values are about 1.4$\sim$1.71. In particular, the best fit sits around $E_c = 1000$~GeV. On the other hand, the $\tau$ final state is found to be always dominant over the three leptonic channels for all of the considered models, while the lifetime is all predicted to be around $1\times 10^{26}$~s. But, according to Refs.~\cite{Cirelli:2012ut,Papucci:2009gd}, all the single-component DM models are ruled out by the diffuse $\gamma$-ray measurements due to the too short lifetimes.  

\begin{table}[ht]
\caption{Parameters leading to the minimal values of $\chi^2$ with the cutoffs of the single DM being 800, 1000, 1200, 1500 and 2000~GeV, respectively.}
{\begin{tabular}{@{}cc|cc|ccc|c|cc@{}} \toprule
$E_{c}({\rm GeV})$ & $m({\rm GeV})$ & $\kappa_1$ & $\kappa_2$ & $\epsilon^{e}$& $\epsilon^{\mu}$  & $\epsilon^{\tau}$
 & $\tau(10^{26}{\rm s})$ & $\chi^2_{\rm min}$ & $\chi^2_{\rm min}/{\rm d.o.f.}$
\\
\colrule
800 & 1654 & 1.01  & 1.72  & 0.07 & 0 & 0.93  & 1.20 & 128 & 1.71 \\
1000 & 2044 & 1.00 & 1.73 & 0.06 & 0 & 0.94 & 1.05 & 103 & 1.37 \\
1200 & 2437 & 0.99 & 1.74 & 0.03 & 0 & 0.97 & 0.93 & 106 & 1.42 \\
1500 & 3030 & 0.99 & 1.74 & 0 & 0 & 1.0  & 0.80 & 125 & 1.67 \\
2000 & 4022 & 0.97 & 1.75 & 0 & 0 & 1.0 & 0.69 & 192 & 2.56 \\
\botrule
\end{tabular}\label{tab1DM_DAMPE}}
\end{table}

\subsection{Double-Component Dark Matter}
\begin{figure}
\centering
\includegraphics[width=0.34\textwidth, angle =-90]{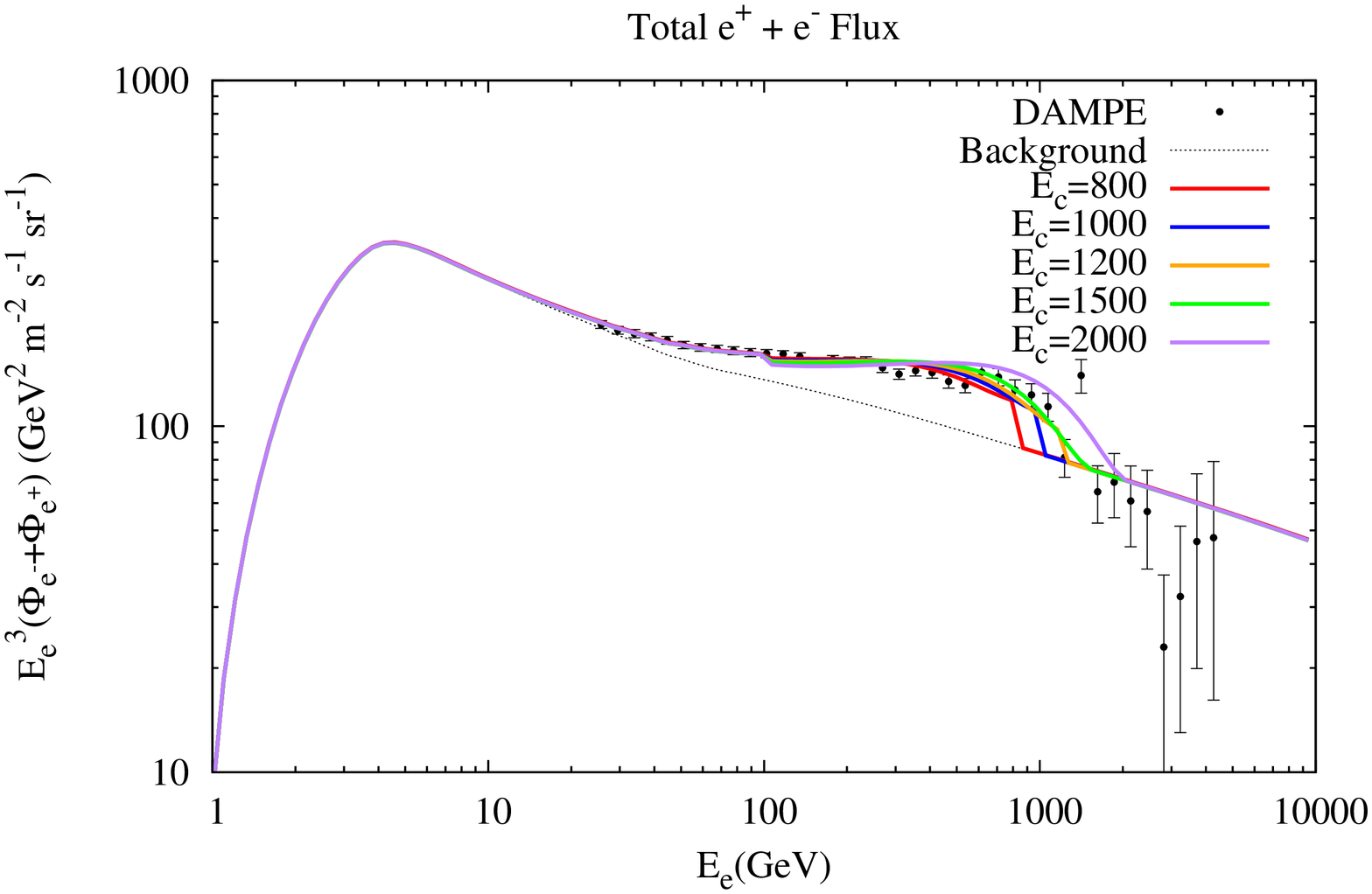}
\includegraphics[width=0.34\textwidth, angle =-90]{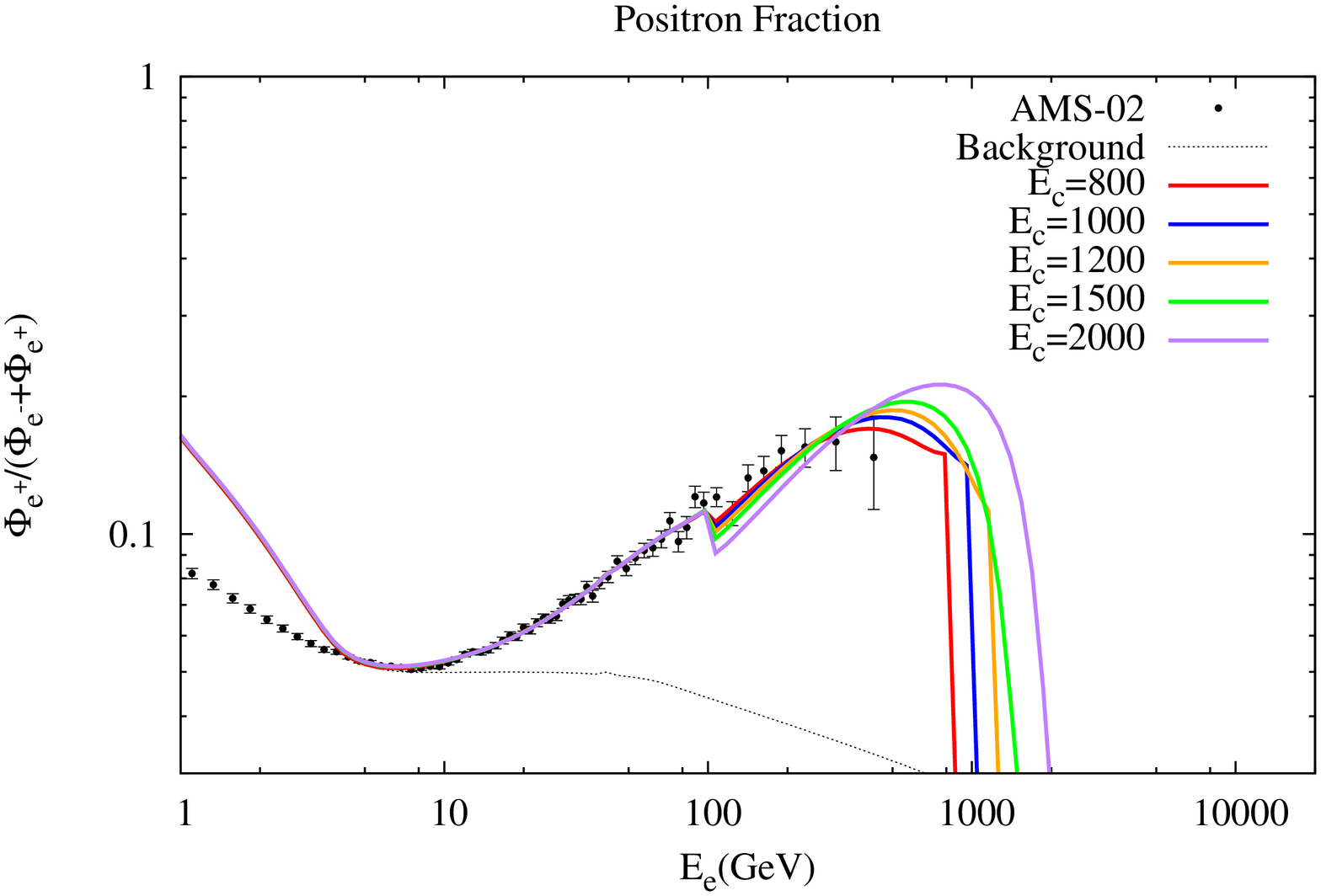}
\caption{The best-fit total $e^+ + e^-$ flux (left panel) and positron fraction (right panel) for double-component DM models with the data from the DAMPE total $e^+ + e^-$ flux and the AMS-02 positron fraction. In each plot, the solid red, blue, orange, green and purple lines (from left to right) correspond to the cases with the heavy DM electron cutoffs to be $E_{cH}=800$, 1000, 1200, 1500 and 2000 GeV, respectively, while the dotted black line represents the common background $e^+/e^-$ spectra.
}\label{Fig2DM_DAMPE}
\end{figure}

Finally, we show the fitting results for the double-component DM models in Table~\ref{tab2DM_DAMPE} and Fig.~\ref{Fig2DM_DAMPE} for $E_{cH} = 800$, 1000, 1200, 1500 and 2000~GeV. From the minimum values of $\chi_{\rm min}^2$, it is seen that 
the model can be reasonably fitted with $\chi^2_{\rm min}/{\rm d.o.f.} \approx 1.5$ when the heavy DM cutoff $E_{cH}$ is taken to be around $1000$~GeV, even though the fit is not so good as those for the corresponding double-component DM models with the CALET+AMS-02 datasets. As for the lepton flavor dependence, we find that both DM decays favor the $\mu$-channel for this best-fit case. Moreover, compared with the single-component DM counterpart, the fit becomes a little worse for the two-component cases since the $\chi^2_{\rm min}$'s  obtained here become larger. The reason lies in the fact that the DAMPE data seems to prefer the $\tau$ final states over other charged leptons, even for the high-energy part of its $e^+ + e^-$ spectrum, which is evident from the flavor preference in Table~\ref{tab1DM_DAMPE} for the single-component DM fits. However, this channel is closed for the two-component DM models in order to avoid the strong diffuse $\gamma$-ray constraint. It is this model assumption that leads to the degradation of the fits here. 



\begin{table}[ht]
\caption{Parameters leading to the minimal values of $\chi^2$ for double-component DM models with the cutoffs of the heavy DM being 800, 1000, 1200, 1500, 2000~GeV, respectively.}
{\begin{tabular}{@{}cc|cc|c|c|cc|cc@{}} \toprule
$E_{cH}({\rm GeV})$ & $m_H(\rm GeV)$ & $\kappa_1$ & $\kappa_2$ & $\epsilon^{e,\mu,\tau}_{L}$& $\epsilon^{e,\mu}_{H}$  & $\tau_{L}(10^{26}{\rm s})$
 & $\tau_{H}(10^{26}{\rm s})$ & $\chi^2_{\rm min}$ & $\chi^2_{\rm min}/{\rm d.o.f.}$
\\
\colrule
800 & 1654 & 1.01 & 1.67 & 0.11, 0.40, 0.49 & 0.17, 0.83 & 2.54 & 1.32 & 134 & 1.86 \\
1000 & 2044 & 1.00 & 1.69 & 0.20, 0.76, 0.04 & 0.13, 0.87 & 3.47 & 1.18 & 110 & 1.53\\
1200 & 2437 & 1.00 & 1.69 & 0.23, 0.77, 0 & 0.07, 0.93 & 3.45 & 1.05 & 112 & 1.56 \\
1500 & 3030 & 1.00 & 1.69 & 0.26, 0.74, 0 & 0, 1.0 &  3.31 & 0.92 & 127 & 1.76\\
2000 & 4022 & 1.00 & 1.69 & 0.32, 0.68, 0 & 0, 1.0 & 3.14 & 0.84 & 180 & 2.50 \\
\botrule
\end{tabular}\label{tab2DM_DAMPE}}
\end{table}

\section{Conclusions and Discussions}
\label{s5}
In the light of the recent release of the CALET~\cite{CALET} and DAMPE~\cite{DAMPE} total $e^+ + e^-$ flux spectra,
we have revisited our earlier multi-component decaying DM proposal~\cite{2DM_1, 2DM_2, 2DM_3, 2DM_4, 2DM_5} to explain the $e^+/e^-$ excesses indicated by the anomalous increase in the positron fraction spectra measured by PAMELA, Fermi-LAT and AMS-02.
In particular, we have tried to fit the datasets of the latest AMS-02 positron fraction spectrum~\cite{AMSf} combined with the
total $e^+ + e^-$ flux of
either CALET~\cite{CALET} or DAMPE~\cite{DAMPE}.

For the combined data from AMS-02 and CALET experiments, both the single and double-component DM models are found to fit the data very well. In the single-component DM cases, the three leptonic decay channels are open for the DM particle. As a result, the DM with its cutoff $E_{c}<1$~TeV is favored and the $\tau$-channel is found to be dominant, which has been excluded by the diffuse $\gamma$-ray observations by Fermi-LAT~\cite{FermiLAT_Gamma,Cirelli:2012ut,Papucci:2009gd}. As for the two-component DM models, the light DM particle with its energy cutoff $E_{cL} = 100$~GeV, motivated by the substructure seen in the positron fraction, is allowed to decay into all lepton channels, while the heavy DM particle decays only into $e^\pm$ and $\mu^\pm$ in order to avoid the strong diffuse $\gamma$-ray constraints. It is found that the best-fit model predicts the heavy DM energy cutoff to be $E_{cH}=1000$~GeV, and the $\tau$($\mu$)-channel dominates the light (heavy) DM decays.

As for fits with the AMS-02 positron fraction and the DAMPE total $e^+ + e^-$ flux data, we take the same model setups and diffusion coefficients for both single-component and double-component DM models as those in fits with the AMS-02+CALET dataset. However, we find that the final fitting results depend on the shape of the primary electron spectrum. Concretely, if we turn the slope index $\gamma_{e3}$ for the primary electron high-energy part to be around 2.5, then the fittings with the AMS-02+DAMPE dataset can be greatly improved. In the present work, we show the fitting results with $\gamma_{e3} = 2.5$. It turns out that both the single- and double-component DM models can provide reasonable fits to the combined AMS-02+DAMPE data, even though the fitting quality is not as good as that with the AMS-02+CALET data. For the single-component DM case, the best fit indicates that the DM energy cutoff should be around $1000$~GeV and the dominant decay channel is the $\tau$ final state, which has already been excluded by the diffuse $\gamma$-ray constraint. On the other hand, the double-component DM models do not suffer from this problem as we turn off the $\tau$-channel for the heavy DM decays. In this case, the best-fitted heavy DM energy cutoff is still $E_{cH} = 1000$~GeV and both DM components prefer the $\mu$-channel as their main decay mode. Note that, as shown in Refs.~\cite{Yuan:2017ysv,Ding:2020wyk}, the fitting can be further improved if we add to the primary electron source spectrum with an exponential cutoff factor, which could be explained very naturally in terms of the cooling effects during the high-energy electron propagation in the Milky Way. 
As a result, we find that both the single- and double-component DM models can fit the AMS-02 and CALET data very well, while the fits to the combined data from AMS-02 and DAMPE are not good enough.

Finally, we remark that the above conclusion is based on our simplified multi-component DM model setup by fixing the primary CR spectra parameters and the CR propagation coefficients listed in Table~\ref{parameters}. One possible problem is that our  final results might be biased by such a choice of parameters. In particular, it is well-known that there is a large uncertainty related to the primary electron component and the diffusion coefficients. The fitting quality might change when the related parameters vary in the physically allowed range. Therefore, the proper way to assess the consistency of different experiments within the multicomponent DM scenario is to allow them to vary along with other DM and source parameters so that we can explore a more comprehensive parameter space. {\color{red} In light of the possible improvements in our fittings, we would like to mention that the comparison of the goodness-of-fit among the considered cases should be taken with caution since the full parameter space has not been carefully surveyed.} Another issue related to our analysis is that the conclusions are obtained by the analysis with the simple $\chi^2$ fitting procedure. It was shown in Ref.~\cite{Andrae:2010gh} that a more appropriate way to discuss the model comparison should be based on the corresponding $p$-value. Furthermore, in order to fully assess the fitting quality, we should also show the uncertainties of our fitting results. Especially, it is particularly useful to plot the signal regions on the model parameter space, say on the $m_{\rm DM}$-$\tau$ ($m_{H}$-$\tau_H$) plane for the single (double)-component DM case. However, such a full statistical analysis requires us to sample over the whole parameter space, which is beyond our ability of the numerical scan in face of so many free parameters in our multi-component DM models. Moreover, as mentioned in Sec.~\ref{s1}, the main motivation in the present paper is to see if, following our previous works in Refs.~\cite{2DM_1, 2DM_2, 2DM_3, 2DM_4, 2DM_5}, our simple multicomponent DM setup, including the model parameter choice and the $\chi^2$ fitting method, could give a good fit to the new $e^+ + e^-$ flux data of CALET and DAMPE together with the more precise AMS-02 measurement of the positron fraction.
 The investigation of the full parameter space and the assessment of the consistency of different experimental datasets with more sophisticated statistic methods
 will be performed in our future study.



\section*{Acknowledgments}
CQG and LY are partially supported by National Center for Theoretical Sciences and MoST (MOST-104-2112-M-009-020-MY3
and MoST-107-2119-M-007-013-MY3), and DH by the Chinese Academy of Sciences (CAS) Hundred-Talent
Program. DH is also supported by Fundacao para a Ciencia e a Tecnologia (FCT), within projects \textit{From Higgs Phenomenology to the Unification of Fundamental Interactions} - PTDC/FIS-PAR/31000/2017 and UID/MAT/04106/2019 (CIDMA) and by the National Science Centre (Poland), research project no.~2017/25/B/ST2/00191.


\end{document}